\documentclass[twocolumn,showpacs,showkeys,preprintnumbers,amsmath,amssymb,aps,nofootXinbib,floatfix]{revtex4}
\usepackage{color}
\usepackage{indentfirst}
\usepackage{graphicx}
\usepackage{multirow}
\begin{document}

\title{
Spin scissors or spin-flip?\\
(comment on the paper in arXiv:2102.13580v1 [nucl-th], 2021)
}

\author{ E.B. Balbutsev\email{balbuts@theor.jinr.ru}}
\affiliation{Bogoliubov Laboratory of Theoretical Physics, Joint Institute for Nuclear Research, 141980 Dubna, Russia}

\begin{abstract}

It is explained that the spin scissors and spin-flip are different names of the same physical phenomenon.
The relation between WFM and RPA methods is clarified.

\end{abstract}


\maketitle

\section{Introduction}

The idea of the scissors type motion in deformed nuclei was reported by
R. Hilton in 1976 \cite{Hilt}. He has in mind the scissors like rotational 
oscillations of protons with respect of neutrons. 
The scissors mode is a magnetic dipole excitation with quantum numbers
$K^{\pi}=1^+$. Such excitations were predicted theoretically \cite{Suzuki}
and found experimentally \cite{Bohle} in the energy interval 2-4 MeV. 
It was understood later
that one can imagine similar oscillations of any pair of nucleus 
constituents (spin up and spin down protons and neutrons) 
versus another pair. In such a way two more possible nuclear 
scissors type excitations were predicted \cite{BaMo,BM3S}: the rotational
oscillations of all spin up nucleons with respect of all spin down nucleons
and the rotational oscillations of spin up protons together with spin down 
neutrons versus spin down protons together with spin up neutrons. They were
called spin scissors.

\begin{table}[b!] 
\caption{$M1$ strengths (in $\mu_N^2$) summed in various energy intervals by \cite{Nest} and \cite{BM3S}.
}
\begin{ruledtabular}\begin{tabular}{cccccc}
            & Ref. &    0-2.4 MeV & 2.4-4 MeV & 0-4 MeV &  Exp.   \\    
            \hline            
\multirow{2}{*}{$^{160}$Dy} &  \cite{Nest} & 1.32 & 4.85 & 6.16 & \multirow{2}{*}{2.42} \\                      
                            &  \cite{BM3S} & 1.84 & 3.35 & 5.19 &                       \\   
\multirow{2}{*}{$^{162}$Dy} &  \cite{Nest} & 1.80 & 4.63 & 6.44 & \multirow{2}{*}{3.45} \\                      
                            &  \cite{BM3S} & 1.80 & 3.58 & 5.38 &                       \\                              
\multirow{2}{*}{$^{164}$Dy} &  \cite{Nest} & 2.11 & 3.94 & 6.05 & \multirow{2}{*}{5.52} \\                      
                            &  \cite{BM3S} & 1.76 & 3.80 & 5.56 &                       \\                              
\end{tabular}\end{ruledtabular}\label{tab1}
\end{table}


\begin{figure}[h!]
\includegraphics[width=0.5\columnwidth]{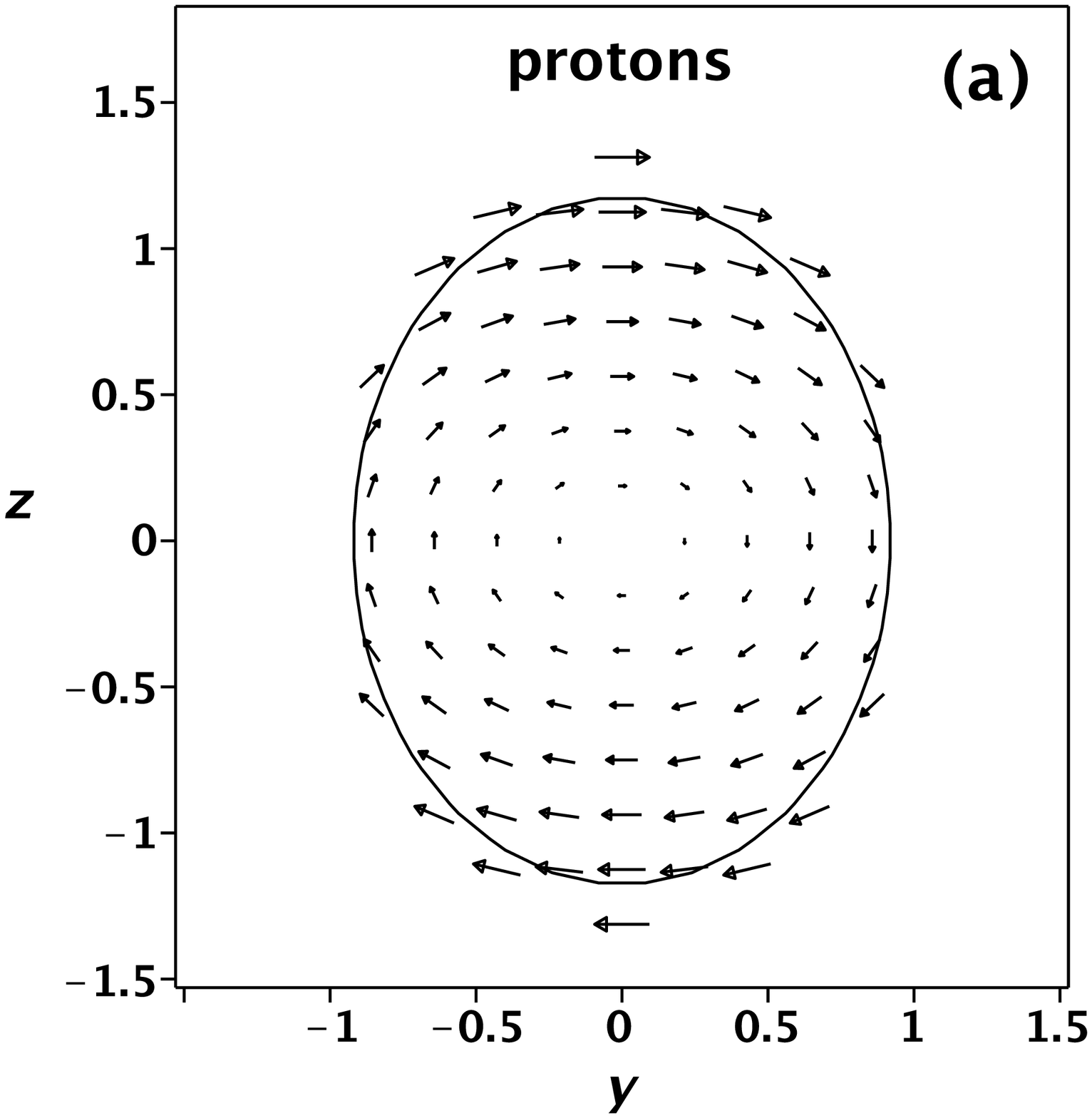}\includegraphics[width=0.5\columnwidth]{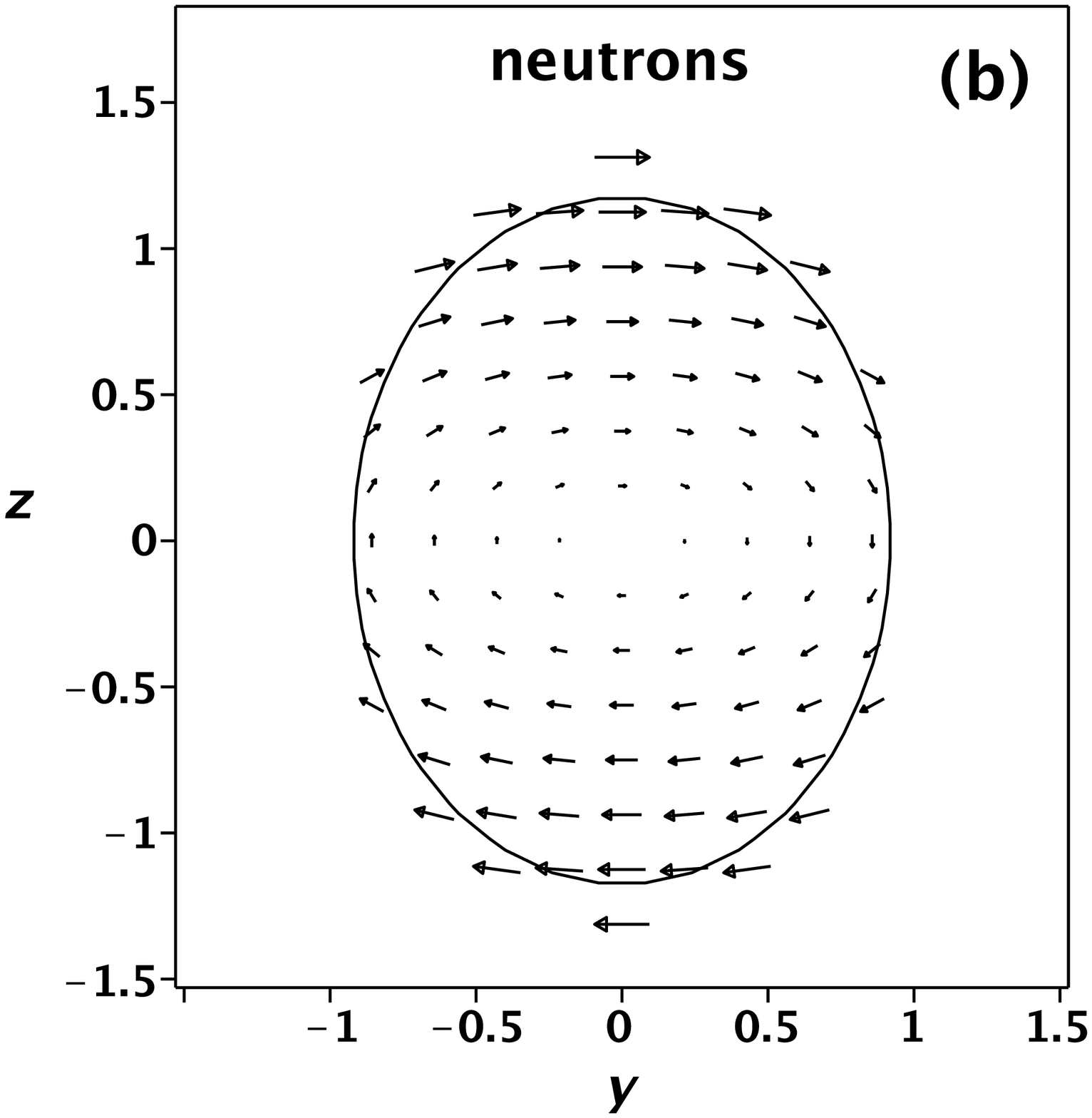}\\
\includegraphics[width=0.5\columnwidth]{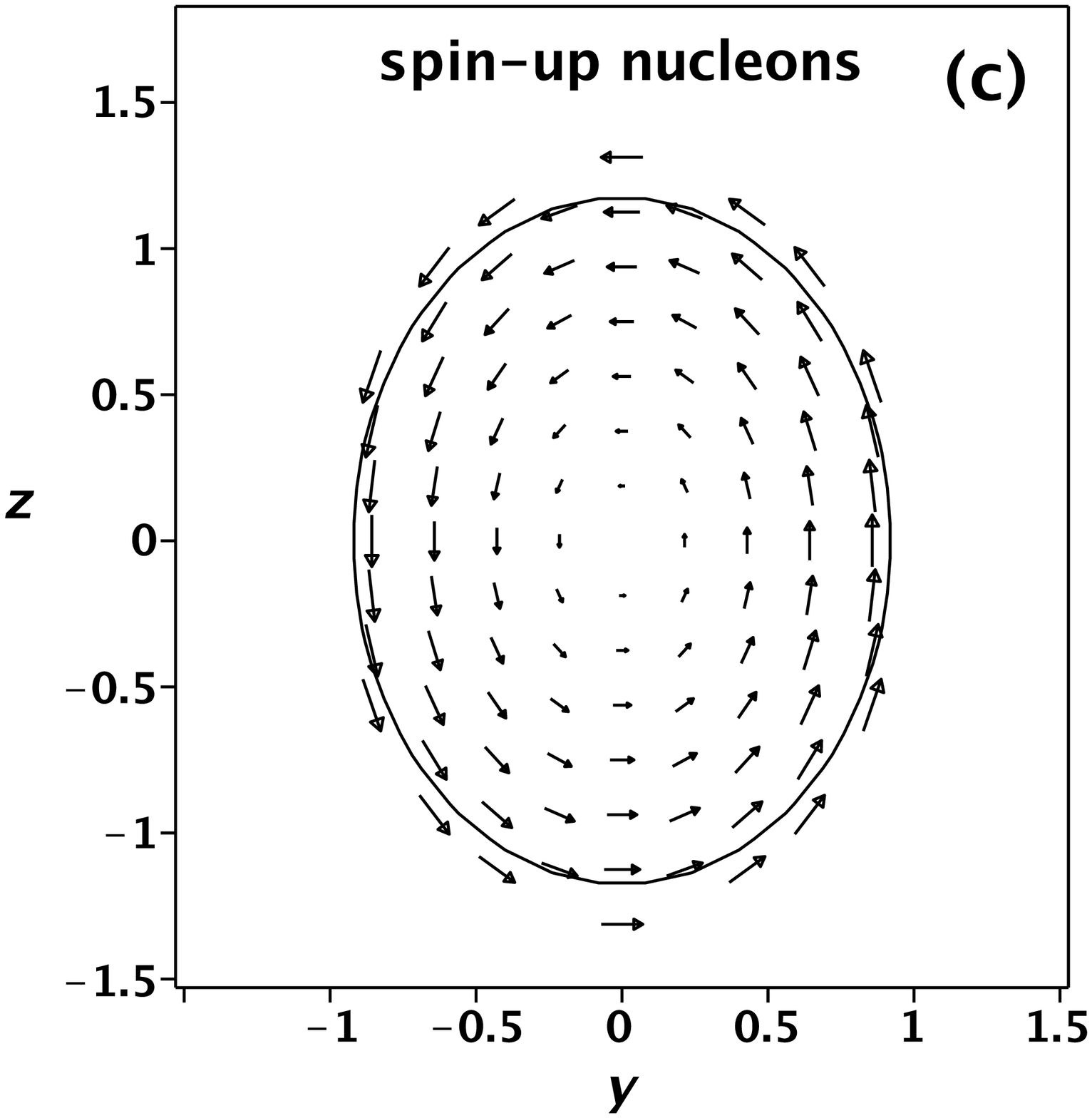}\includegraphics[width=0.5\columnwidth]{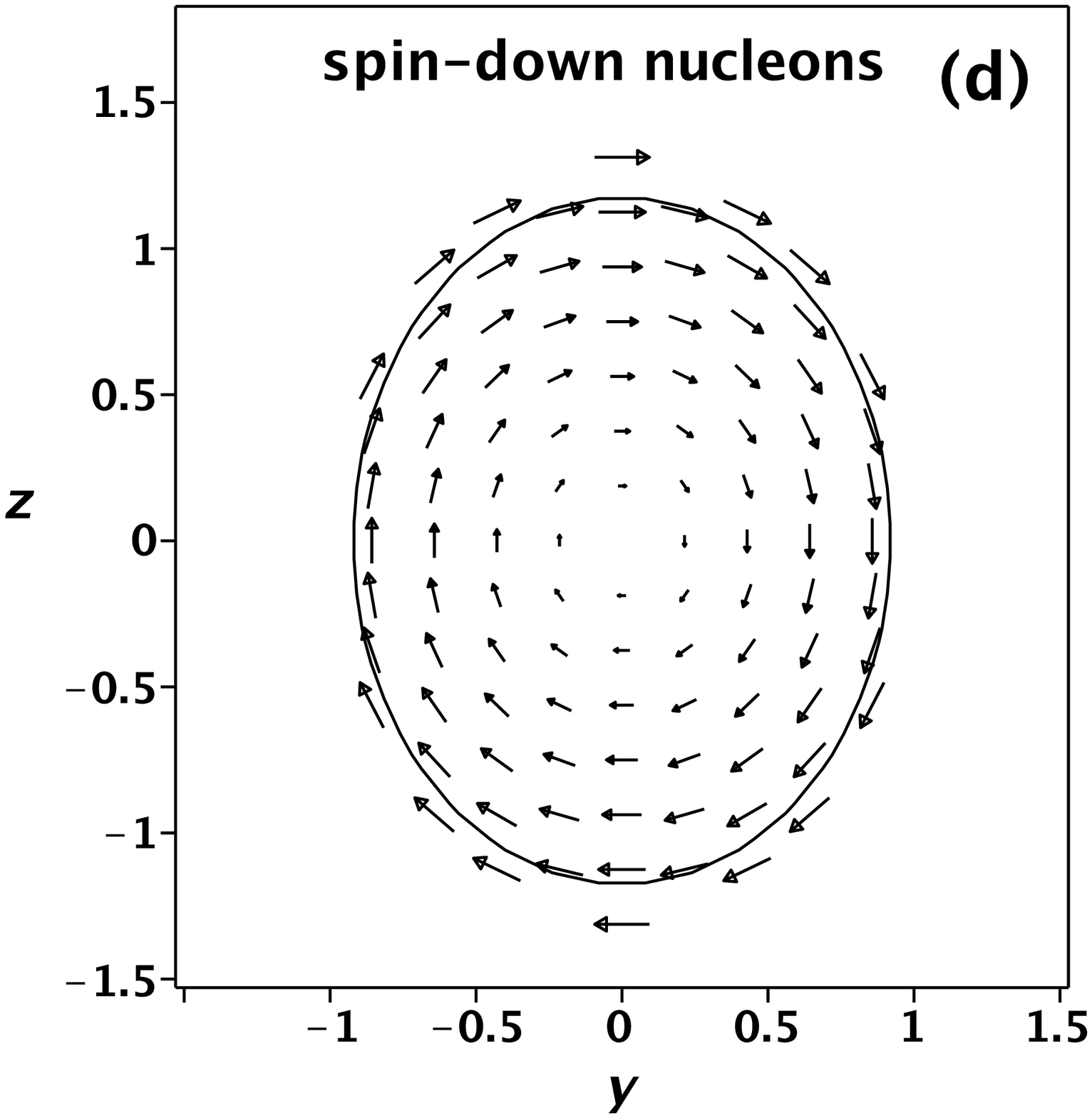}\\
\includegraphics[width=0.5\columnwidth]{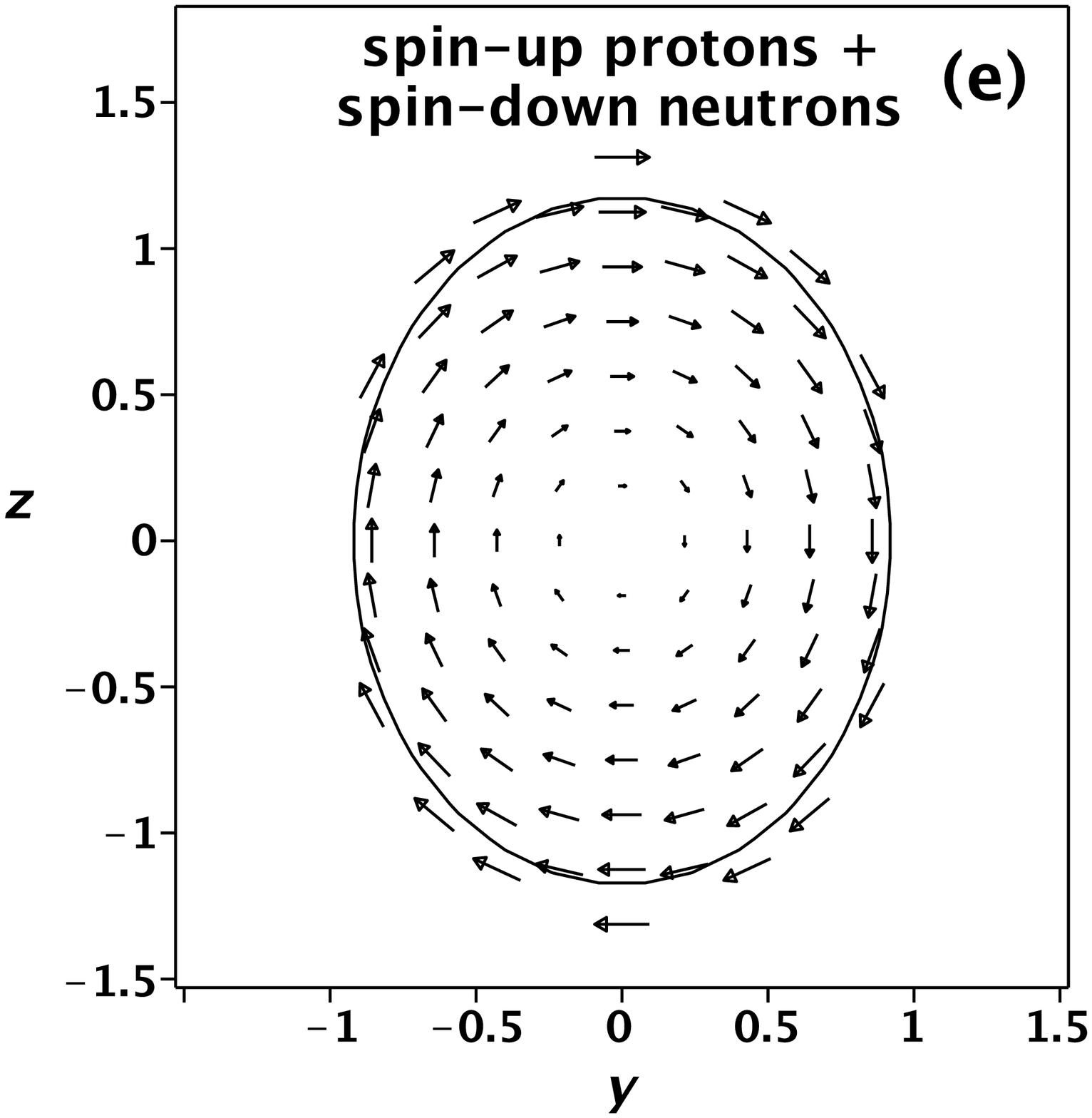}\includegraphics[width=0.5\columnwidth]{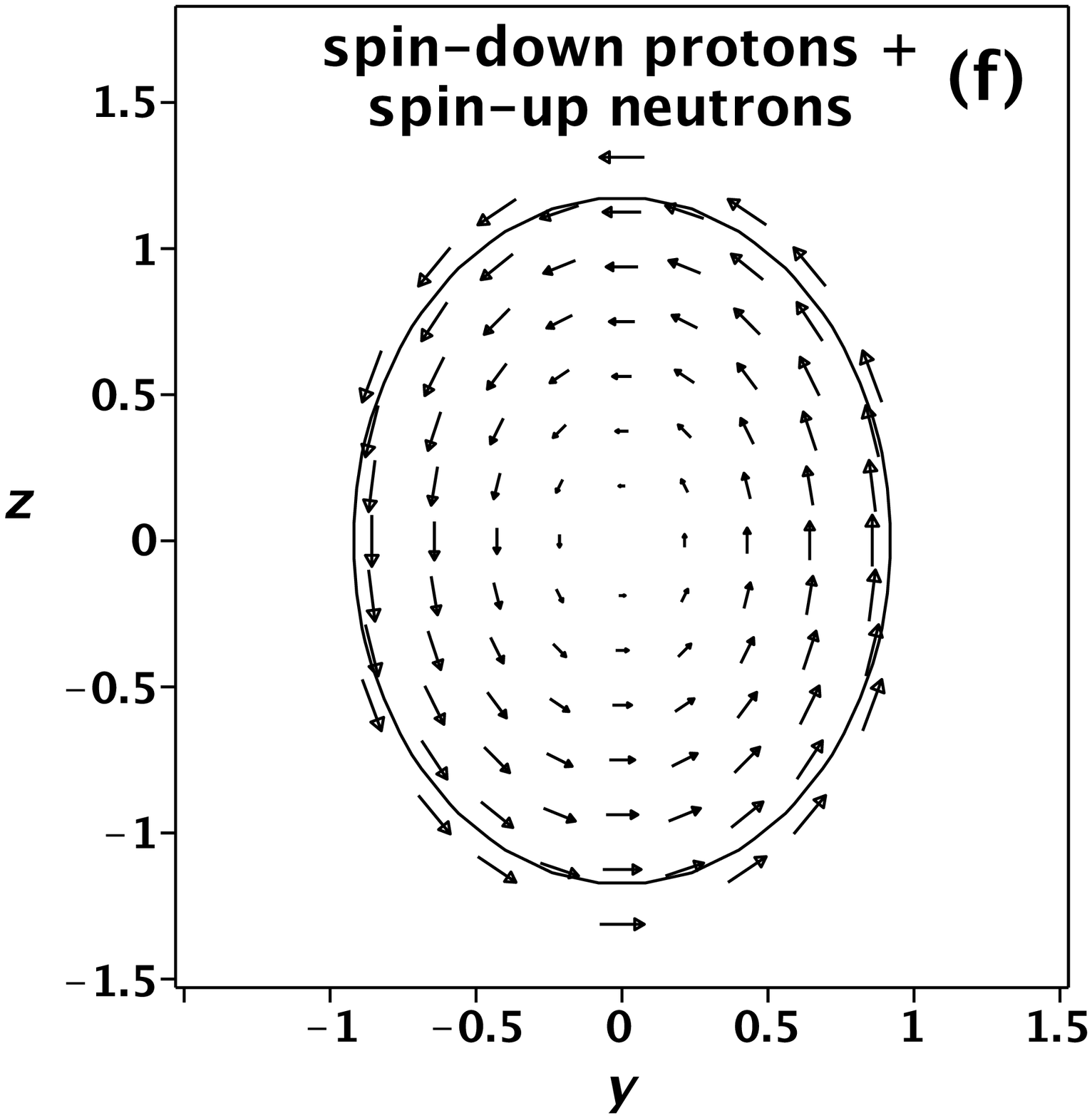}
\caption{The currents $J_{\tau}^{\varsigma}$ in $^{164}$Dy for $E=2.20$ MeV:
$ J^{+}_{\rm p}$~(a), $ J^{+}_{\rm n}$~(b), 
$ J^{\uparrow\uparrow}$~(c), $ J^{\downarrow\downarrow}$~(d),
$ J^{\uparrow\uparrow}_{\rm p}+ J^{\downarrow\downarrow}_{\rm n}$~(e), 
$ J^{\downarrow\downarrow}_{\rm p}+ J^{\uparrow\uparrow}_{\rm n}$~(f).
\mbox{{\textsf y} $=y/R$, {\textsf z} $=z/R$.}}
\label{E1}
\end{figure}

\begin{figure}[t!]
\includegraphics[width=0.5\columnwidth]{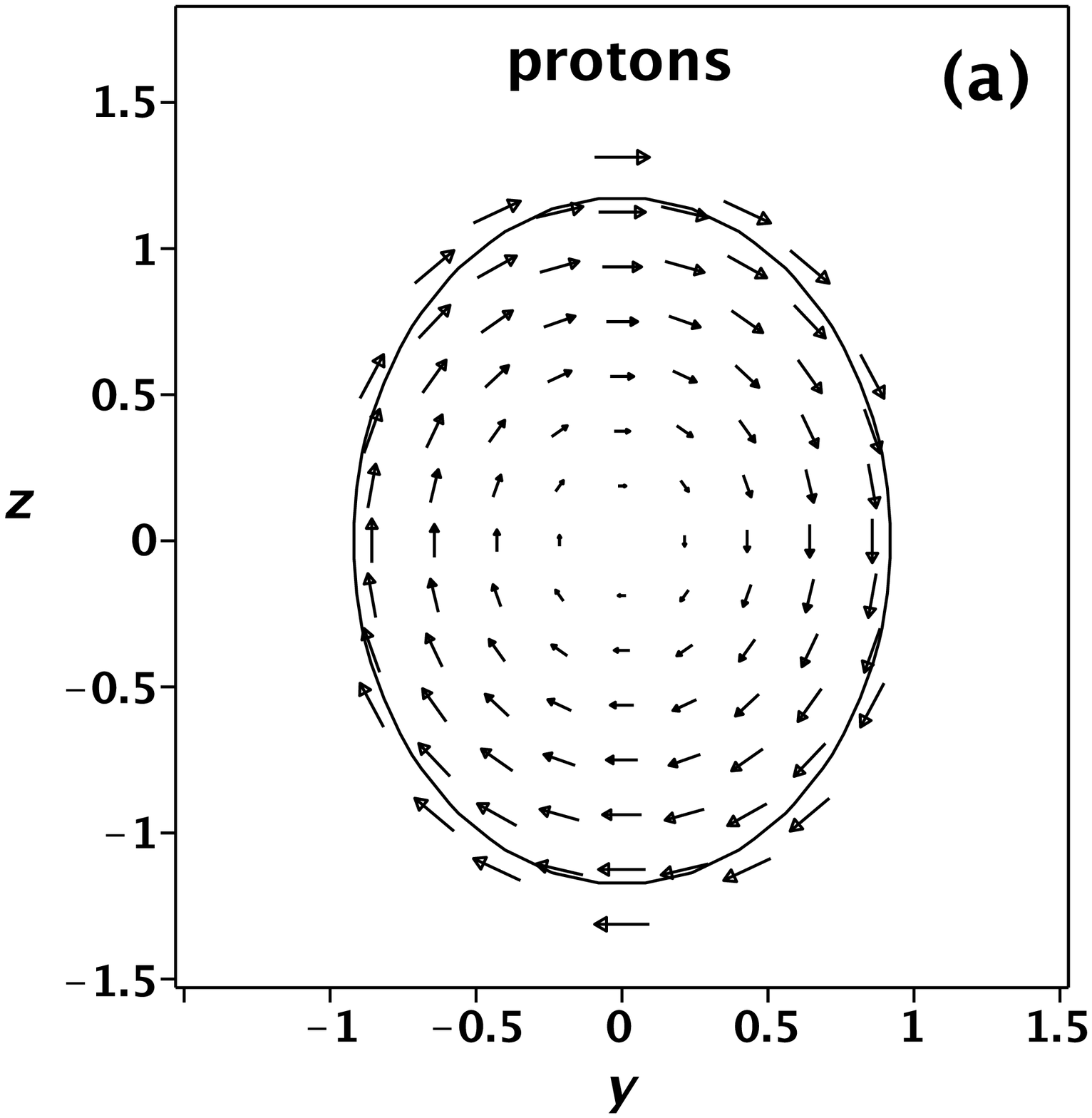}\includegraphics[width=0.5\columnwidth]{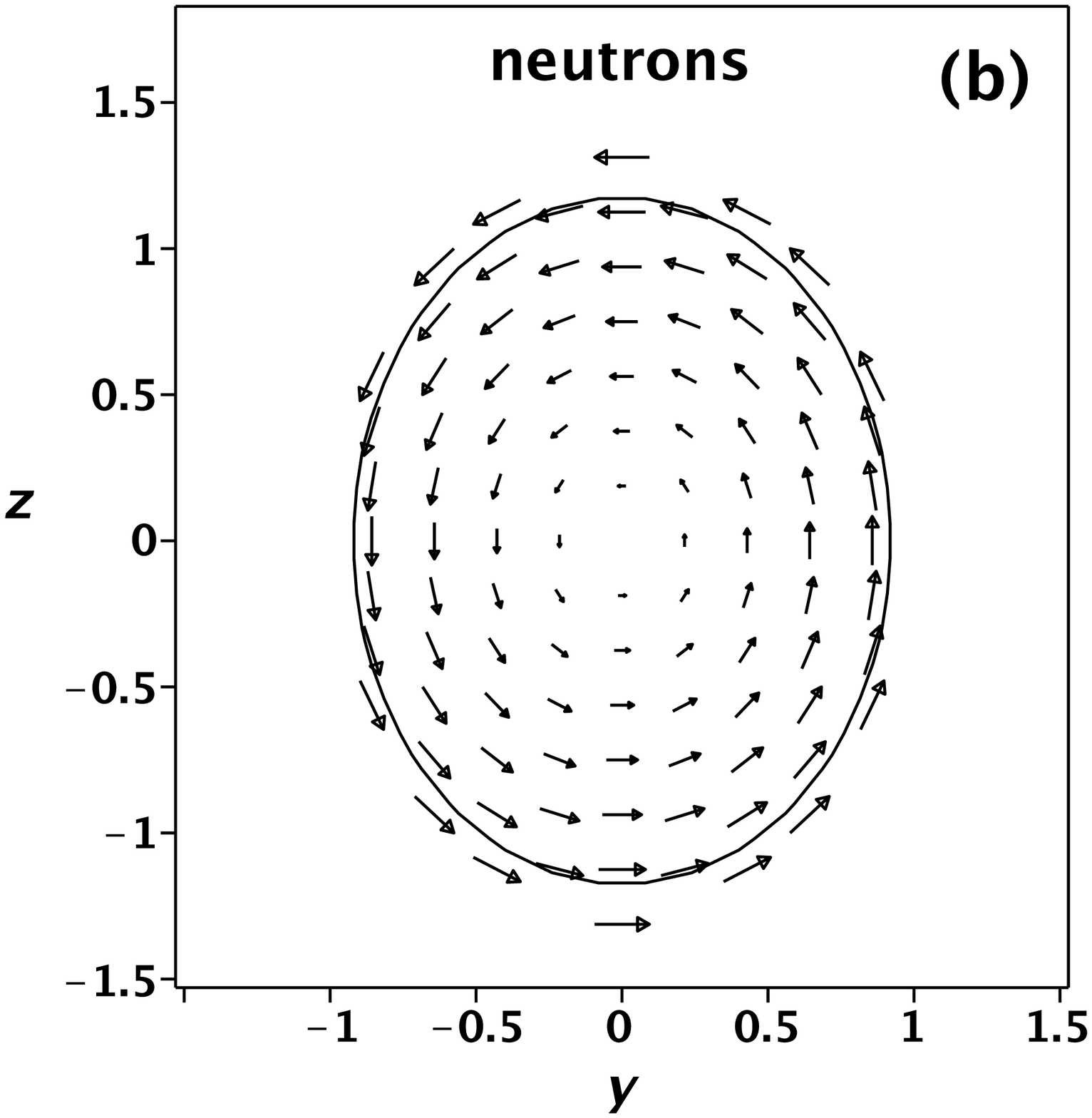}\\
\includegraphics[width=0.5\columnwidth]{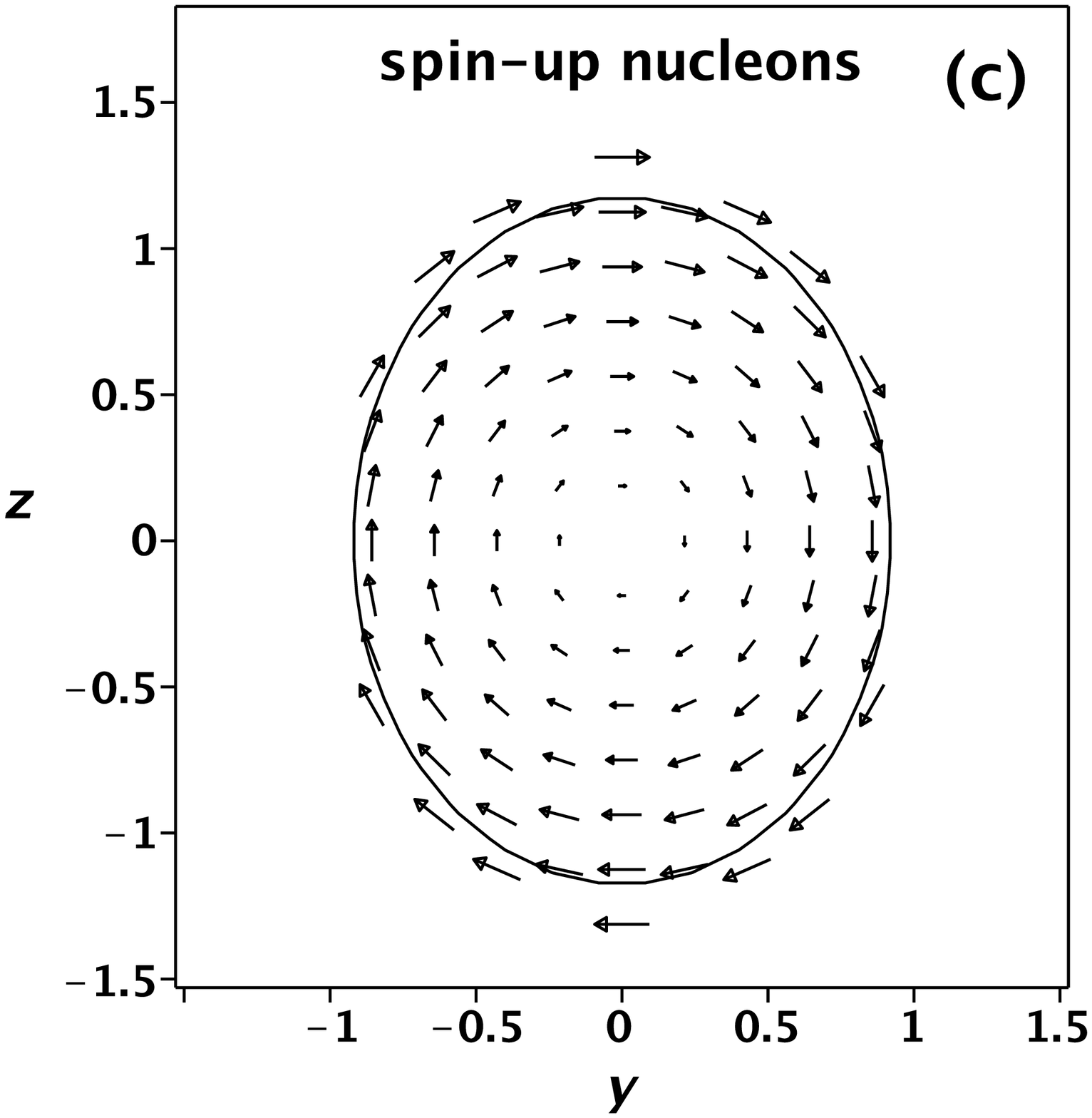}\includegraphics[width=0.5\columnwidth]{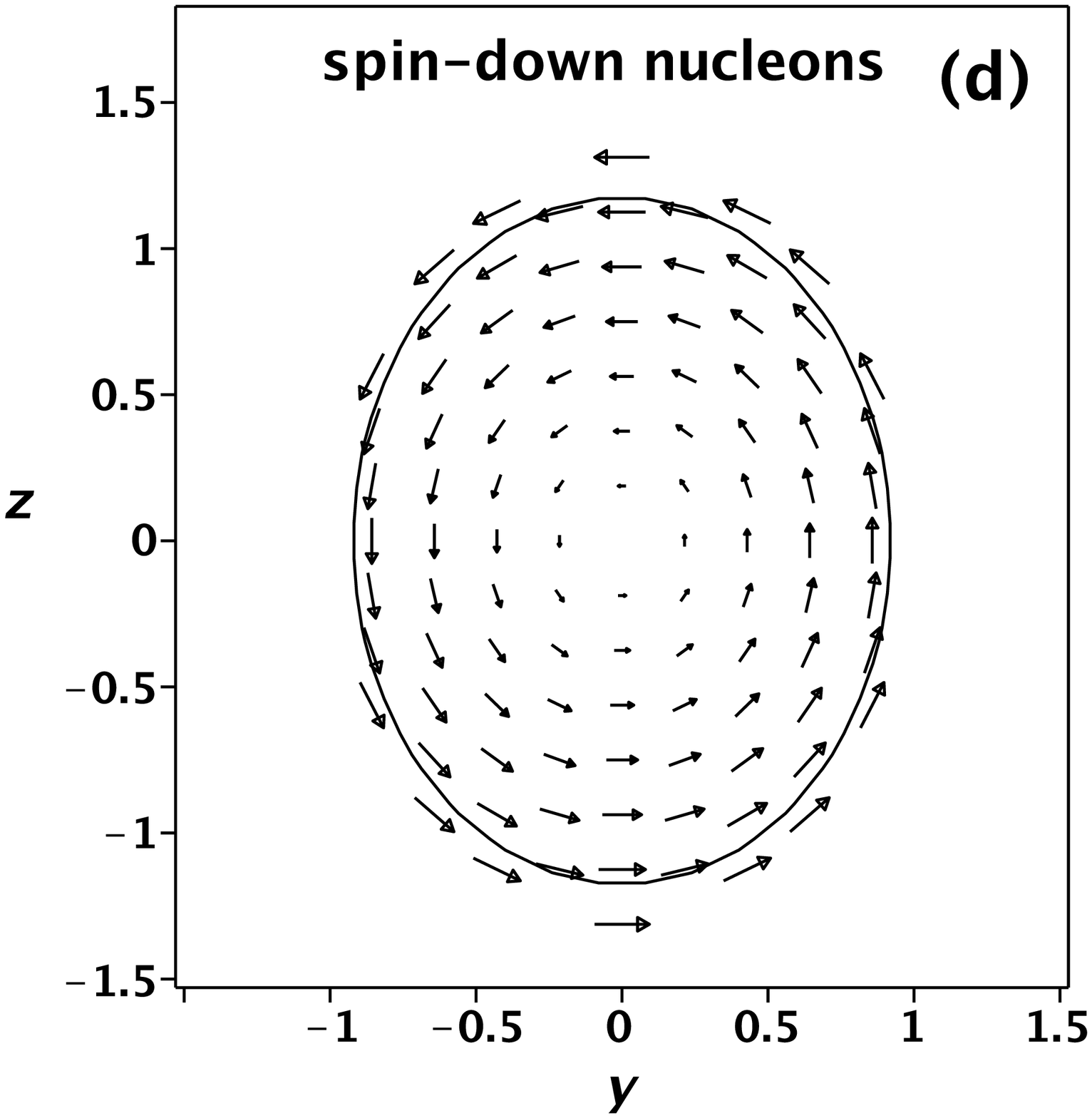}\\
\includegraphics[width=0.5\columnwidth]{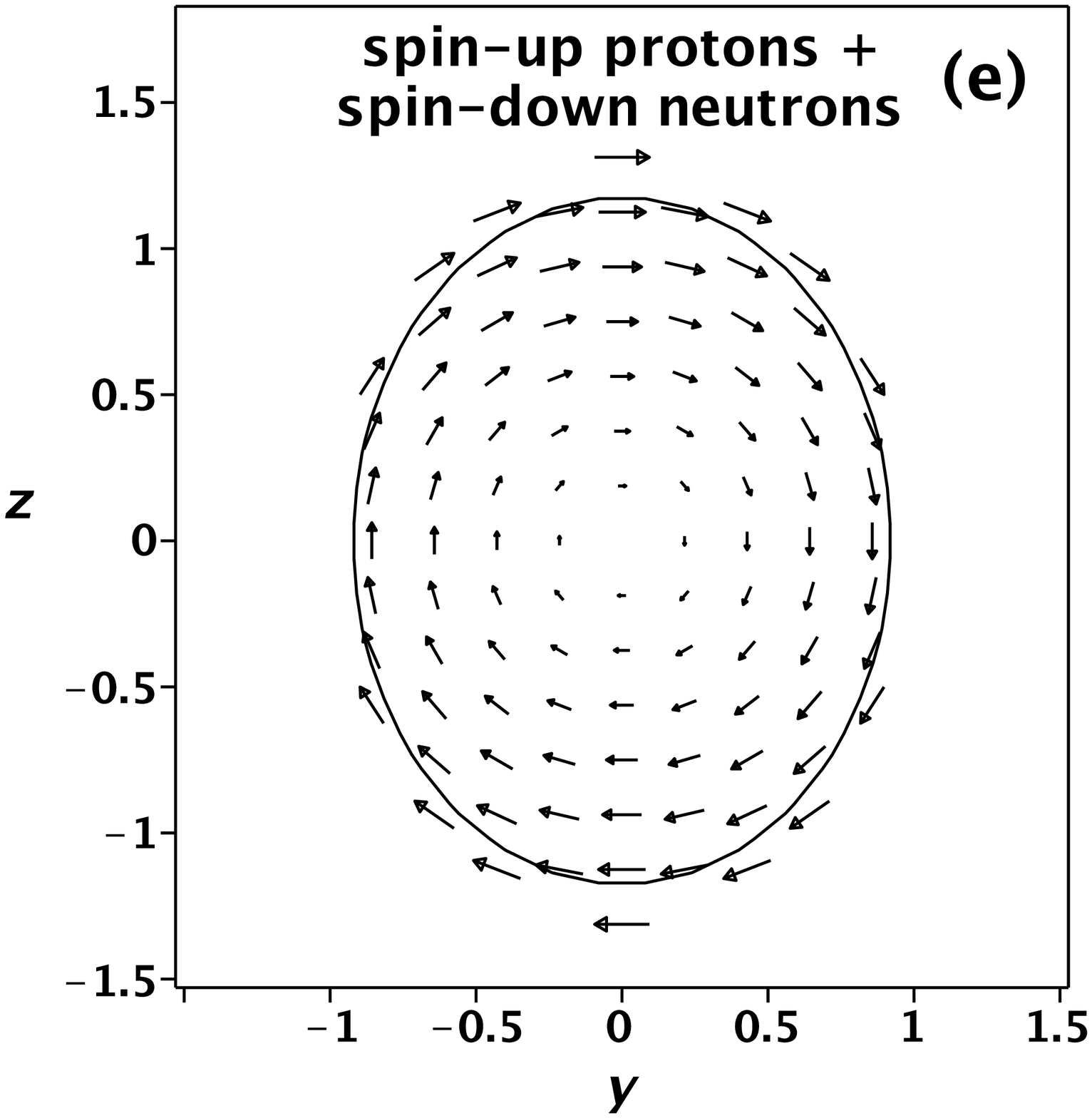}\includegraphics[width=0.5\columnwidth]{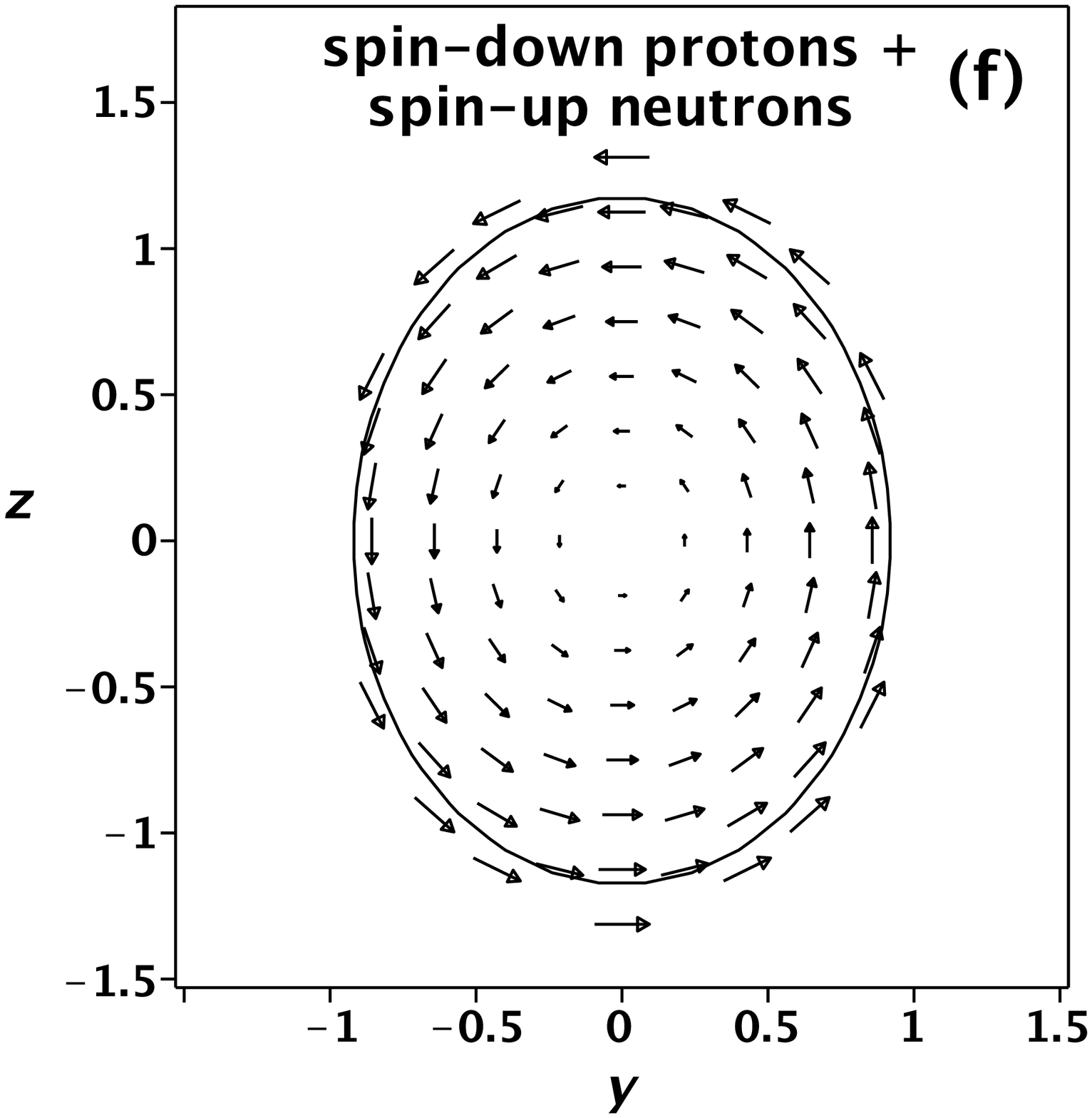}
\caption{The currents $J_{\tau}^{\varsigma}$ in $^{164}$Dy for $E=2.87$ MeV: $ J^{+}_{\rm p}$~(a), $ J^{+}_{\rm n}$~(b), 
$ J^{\uparrow\uparrow}$~(c), $ J^{\downarrow\downarrow}$~(d),
$ J^{\uparrow\uparrow}_{\rm p}+ J^{\downarrow\downarrow}_{\rm n}$~(e), 
$ J^{\downarrow\downarrow}_{\rm p}+ J^{\uparrow\uparrow}_{\rm 
n}$~(f).
\mbox{{\textsf y} $=y/R$, {\textsf z} $=z/R$.}}
\label{E2}
\end{figure}\begin{figure}[t!]
\includegraphics[width=0.5\columnwidth]{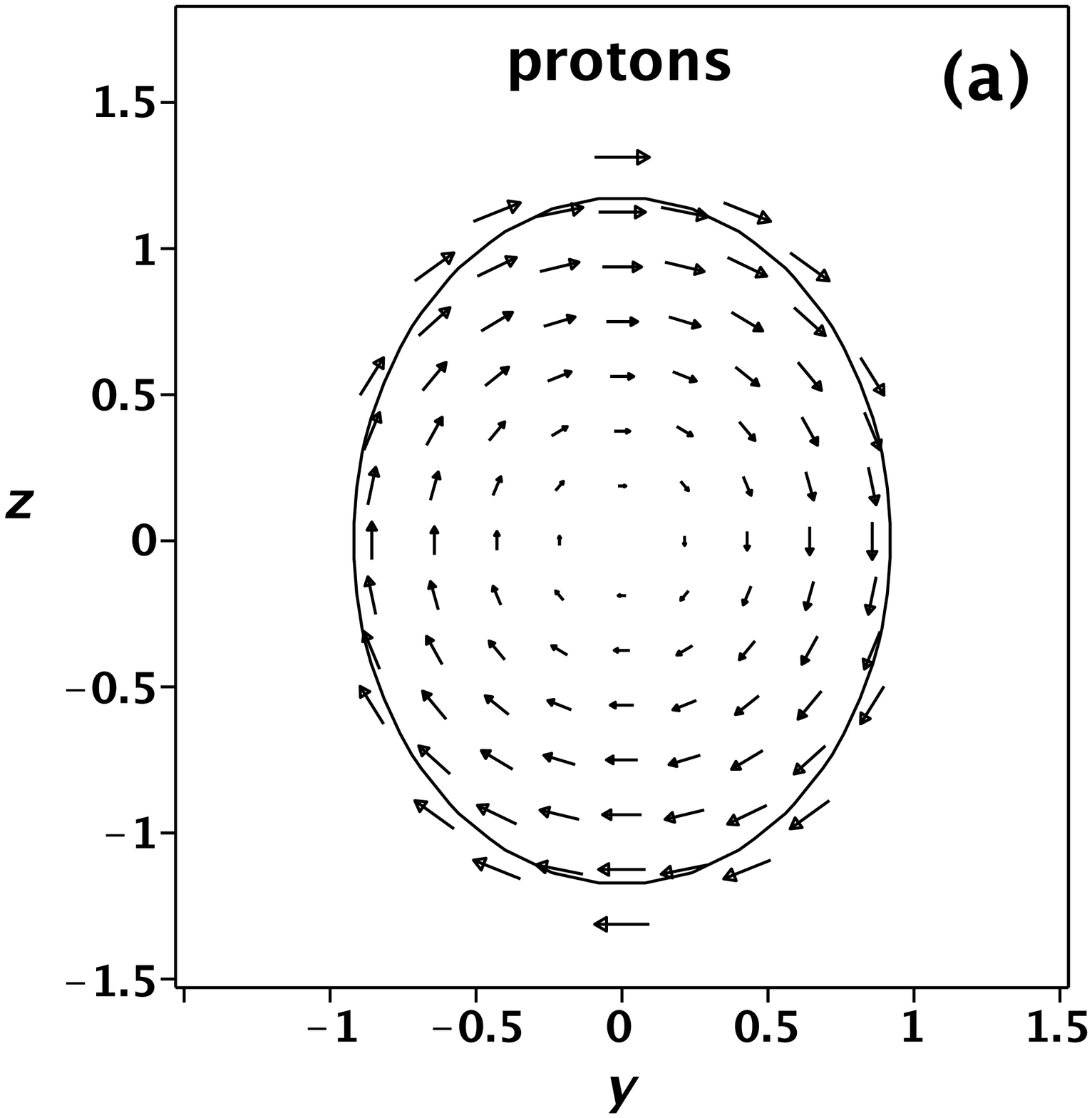}\includegraphics[width=0.5\columnwidth]{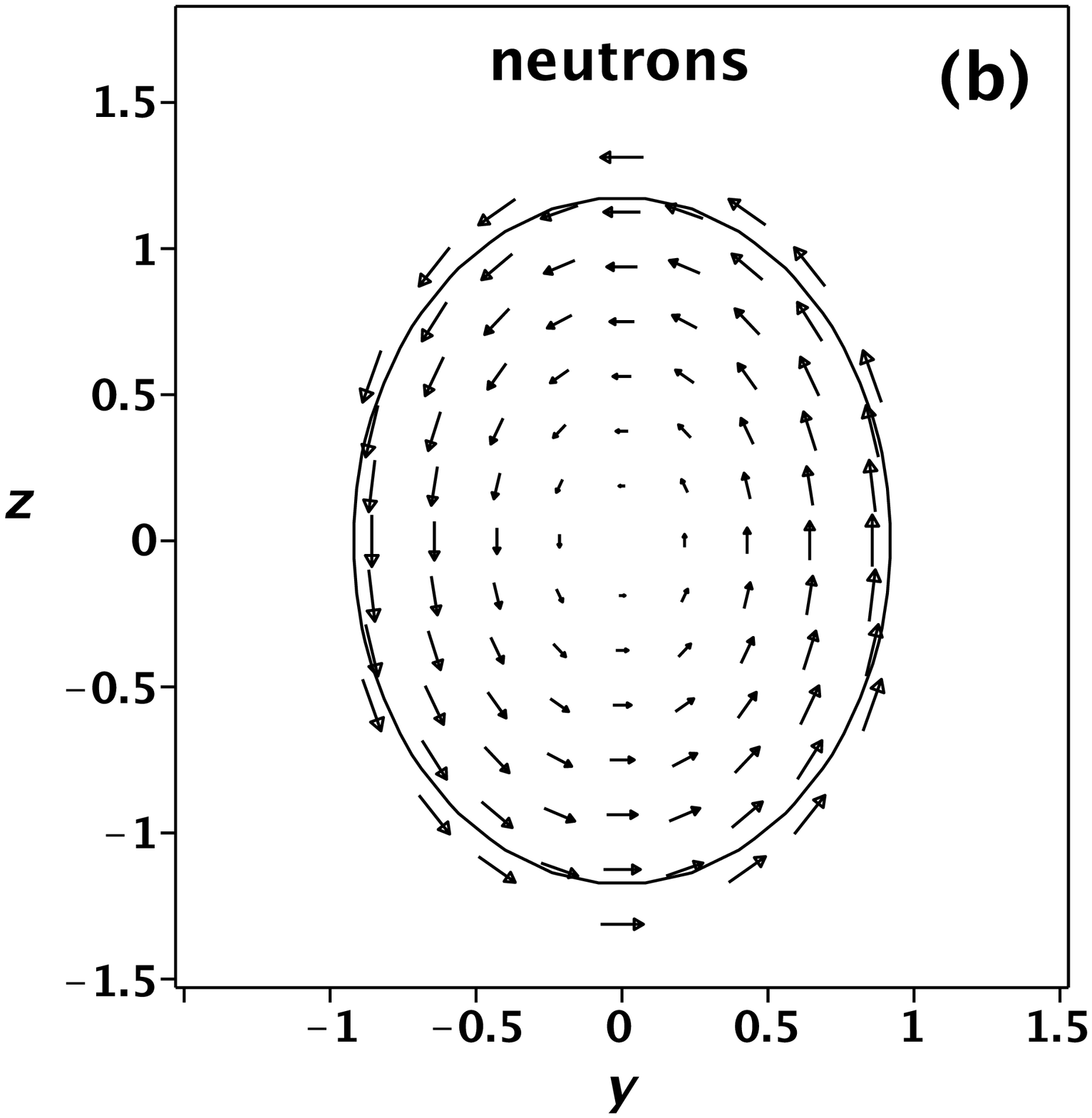}\\
\includegraphics[width=0.5\columnwidth]{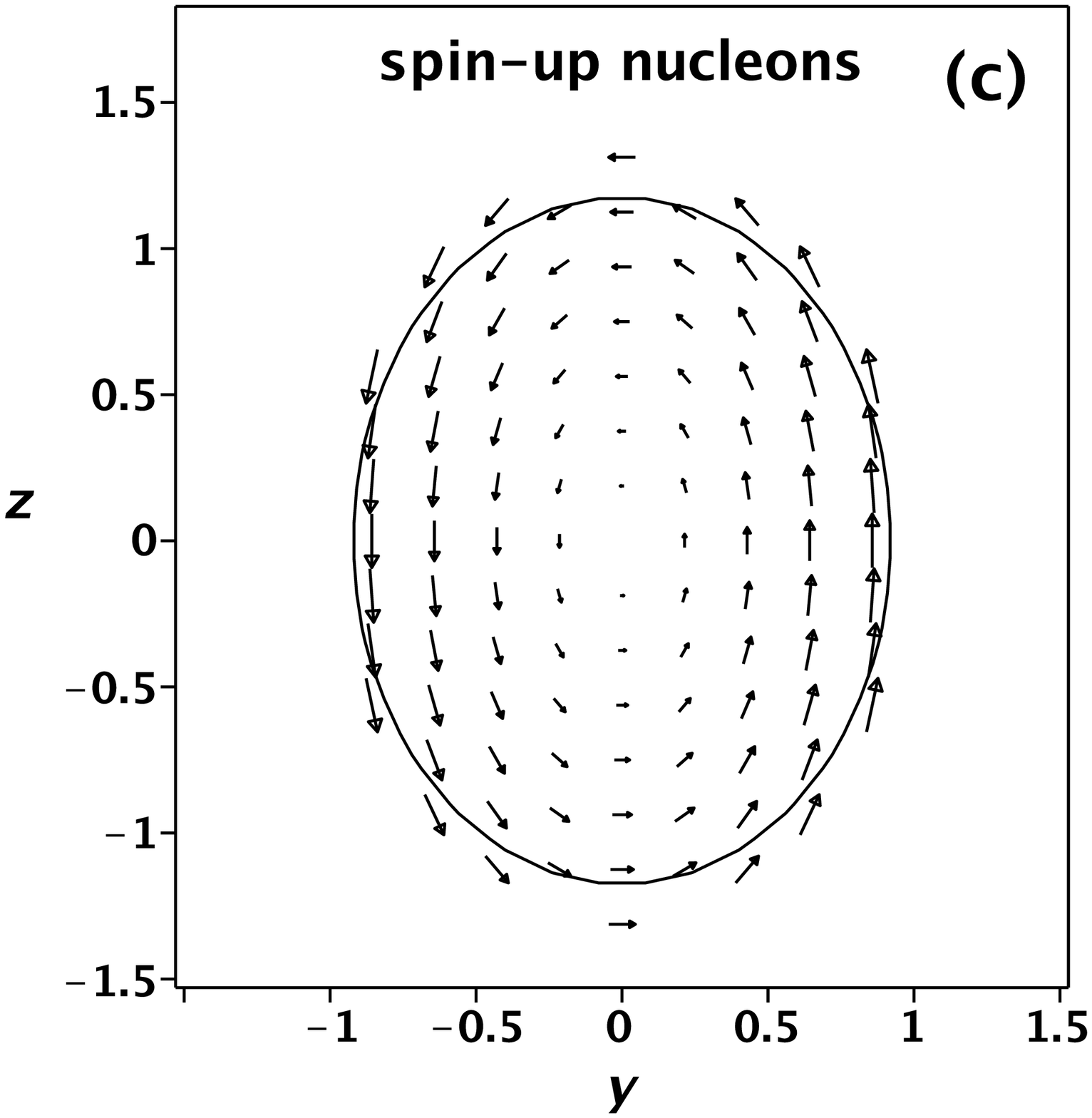}\includegraphics[width=0.5\columnwidth]{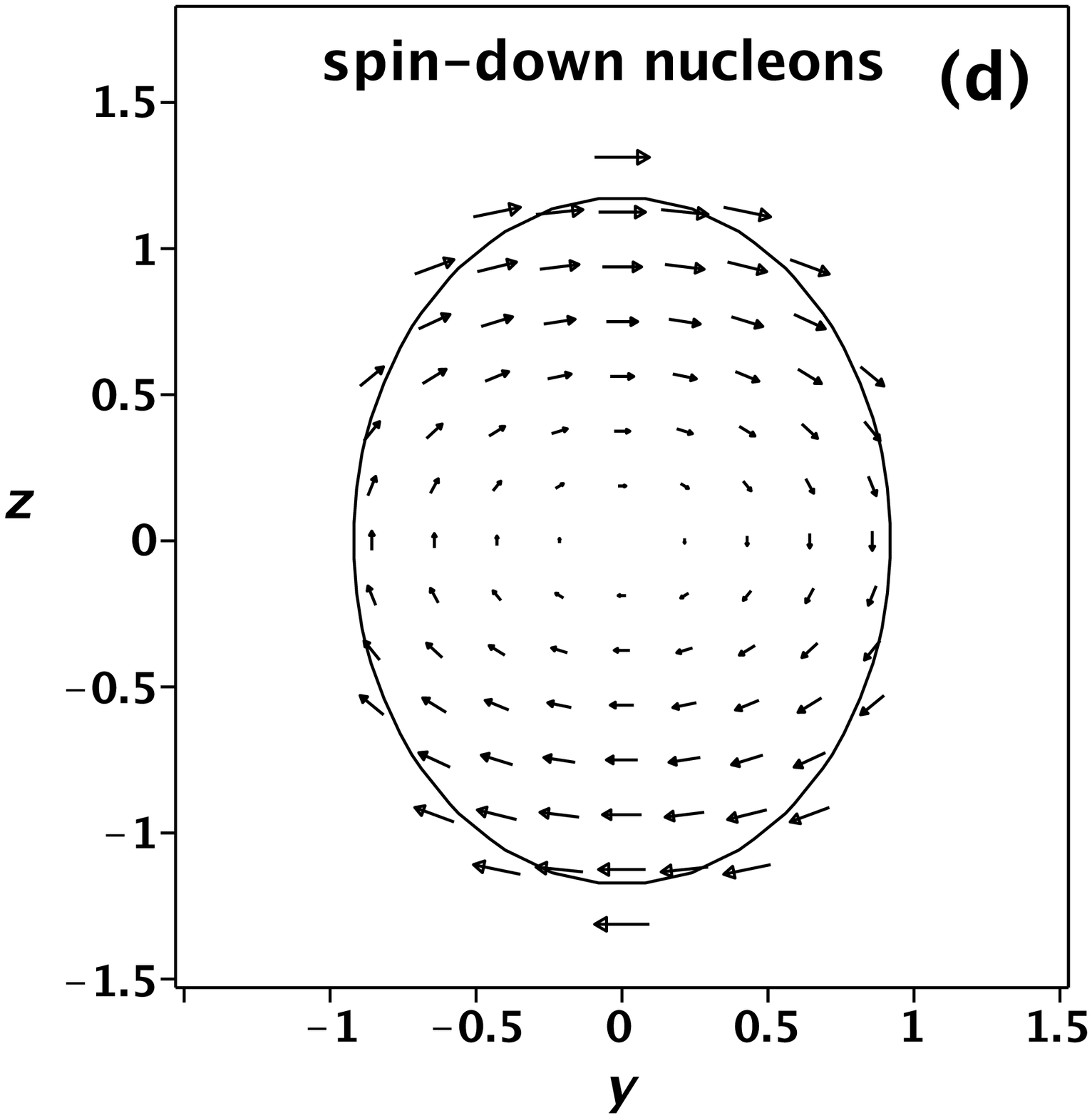}\\
\includegraphics[width=0.5\columnwidth]{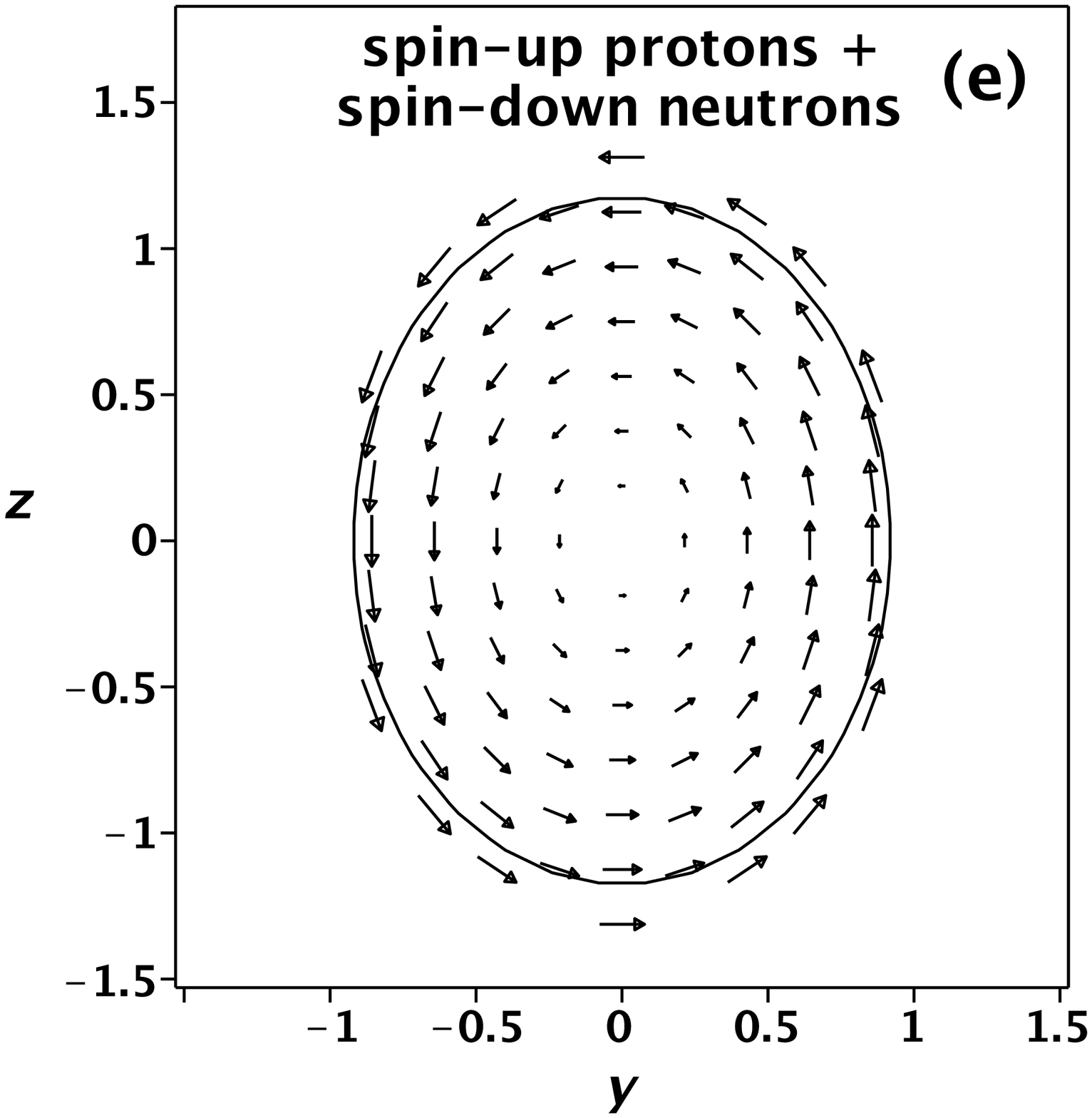}\includegraphics[width=0.5\columnwidth]{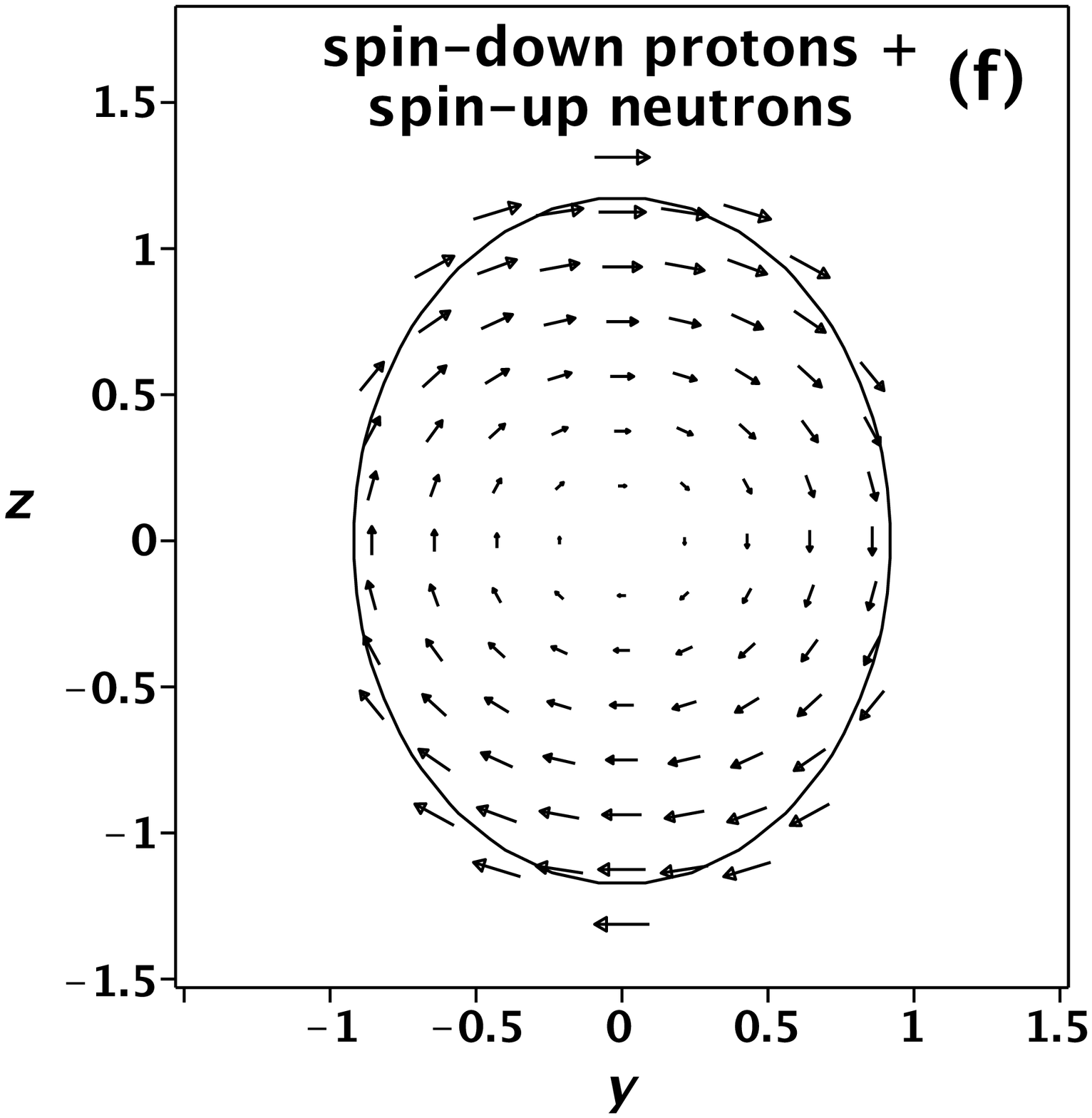}
\caption{The currents $J_{\tau}^{\varsigma}$ in $^{164}$Dy for $E=3.59$ MeV: $ J^{+}_{\rm p}$~(a), $ J^{+}_{\rm n}$~(b), 
$ J^{\uparrow\uparrow}$~(c), $ J^{\downarrow\downarrow}$~(d),
$ J^{\uparrow\uparrow}_{\rm p}+ J^{\downarrow\downarrow}_{\rm n}$~(e), 
$ J^{\downarrow\downarrow}_{\rm p}+ J^{\uparrow\uparrow}_{\rm 
 n}$~(f).
\mbox{{\textsf y} $=y/R$, {\textsf z} $=z/R$.}}
\label{E3}
\end{figure}

\section{Spin-flip}

The low lying $1^+$ states of $^{160,162,164}$Dy and $^{232}$Th were
considered recently in \cite{Nest}.
The calculations were "performed within fully self-consistent 
Quasiparticle Random Phase Approximation (QRPA) with the Skyrme forces SG2, 
SKM* and SVbas". 
The aim of this paper was (as declared by authors) "to scrutinize the
WFM prediction of SSR (Spin Scissors Resonans) from the microscopic 
viewpoint". The summed  $B(M1)$ values of Dy isotopes and $^{232}$Th
turned out rather close to that ones found by WFM method \cite{BM3S}. 
For Dy even the distributions of $M1$ strength over energy regions 
0 -- 2.4 MeV and 2.4 -- 4
MeV are rather similar in the case of SKM$^*$ forces, see Table~\ref{tab1}.
The experimental value of summed  $B(M1)=5.52$ $\mu_N^2$ for $^{164}$Dy 
in Tab. I differs
from that of shown in the Tab. IV of \cite{Nest} ($B(M1)=6.17$ $\mu_N^2$)
because we take into account only the levels \cite{Margraf} with the known 
parity or the branching ratio $R_{expt}< 1$.

The value of summed $B(M1) = 5.77$ $\mu_N^2$ for $^{232}$Th found by WFM is
just in between 4.70 $\mu_N^2$ and 6.69 $\mu_N^2$ found in \cite{Nest} with
different forces (see their Tab. VI). So, the situation with the objective
results is rather good. However, analyzing the results of our paper the 
authors of \cite{Nest} made several questionable and sometimes simply
erroneous statements.

Let us discuss them step by step.

\cite{Nest}, page 2, left column: 

"Because of the scissors nature, these states can exist only in deformed 
nuclei." 

The second part of this phrase is correct only for the conventional 
scissors (see discussion below). Moreover, it is not clear, what the authors
of \cite{Nest} have in mind under "the scissors nature"?

\cite{Nest}, page 2, right column:
 
"So WFM correctly predicts SSR as spin-flip transitions".

Not at all.
The term "spin-flip" belongs to RPA language, not WFM one! WFM and RPA are 
two different approaches describing in this case the same physical 
phenomenon having the same origin -- spin-orbital potential.
 
\cite{Nest}, page 2, right column: 

"Fig. 2 shows that, in contradiction with WFM picture, the deformation is 
not the primary origin of SSR". 

\begin{figure}[t!]
\centering
\includegraphics[width=\columnwidth]{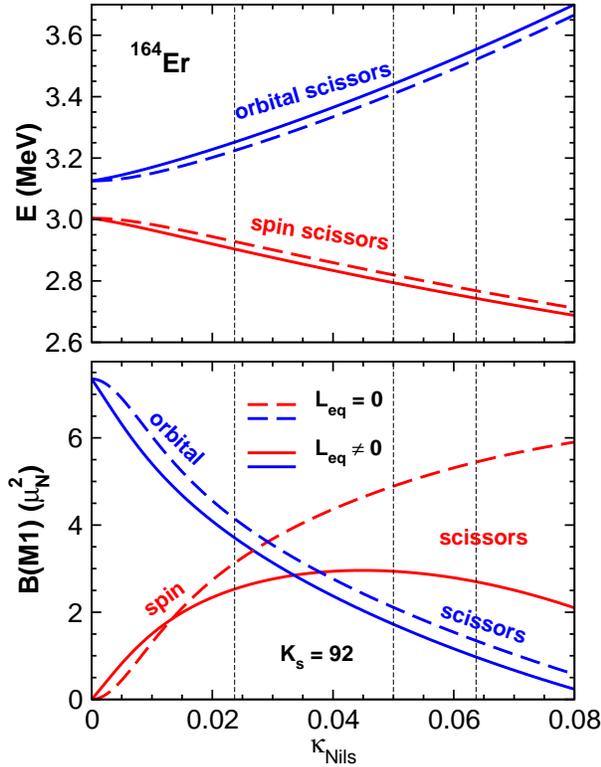}
\caption{The energies $E$ and $B(M1)$-factors as a functions of the spin-orbital strength constant 
$\kappa_{Nils}$. The dashed lines -- calculations without 
$L_{10}^-({\rm eq})$, the solid lines -- $L_{10}^-({\rm eq})$ are taken into account.}
\label{figL}
\end{figure}

Correct finding, but why in contradiction?!
Of course, we consider deformed nuclei, but it does not mean that spin scissors are induced by deformation. We never said anything similar. More
of it, the detailed analysis of various features of this mode was given in
\cite{BaMo} where we have introduced for a first time such type of the 
nuclear collective motion. It was shown there that spin scissors are 
generated by the spin-orbital part of the mean field. The analysis was  continued in the paper \cite{PRC15}, where the strong influence of
antiferromagnetic properties of nuclei on the excitation probability of spin scissors was discovered. The dependence of the energy $E$ and the excitation 
probabilities $B(M1)$ of spin and orbital scissors on the strength of a spin-orbital potential is shown on Fig.~\ref{figL} taken from the paper \cite{PRC15}.

\section{Interpretation, discussion}

\cite{Nest}, page 4, right column:

"Instead, a few states at
E $<$ 2.4 MeV exhibit the spin significant strength... Following prediction
\cite{BaMoPRC1,BM3S}, these states are candidates for SSR". 

It is 
necessary to explain here, that there is the principal difference between 
our definition of the orbital and spin excitations and that of 
the paper \cite{Nest}. Our definitions: 
1) orbital scissors -- rotational oscillations of protons with respect of 
neutrons, 
2) simple spin scissors -- rotational oscillations of all spin up 
nucleons with respect of all spin down nucleons, 
3) complicate spin scissors -- rotational oscillations of spin up protons 
together with spin down neutrons versus spin down protons together with 
spin up neutrons. 

The definition of \cite{Nest} is connected with the contribution of the
proper part of the magnetic dipole operator into $B(M1)$ of the studied 
excitation: it is called orbital excitation, if the contribution of the 
orbital part is bigger than the contribution of the spin part and it is 
called the spin excitation in the opposite situation. In our opinion such 
definition is not very reliable because of the strong (sometimes very 
strong) interference of spin and orbital contributions.

Different definitions lead to the different interpretation of experimental 
data. Let us consider the well known situation with two groups of 1$^+$
states observed in $^{164}$Dy. According to the experimental data from
\cite{Margraf} the summed $M1$ strength above 2.7 MeV is $B(M1)=3.85$ $\mu_N^2$
and below 2.7 MeV it is $B(M1)=1.67$ $\mu_N^2.$ Because of the difficulties
with the determination of the parities of excited dipole states it was
decided in the paper \cite{Pietr95} to refer to OSR (Orbital Scissors Resonance) only the states disposed in the energy interval 
2.7 MeV $< E <$ 3.7 MeV for Z $<$ 68. That is why the lower group was not 
taken into account in this systematics of OSR. In another systematics
\cite{Rich} the energy interval was extended to 2.5 MeV $< E <$ 4 MeV, but
the lower group was omitted again as an exception.

An additional argument to omit the lower group "is the existence of 
low-lying two-quasiparticle excitations around 2.5 MeV as reported from a 
particle transfer experiment \cite{Freem} on the nucleus $^{164}$Dy".

It is necessary to say about one more possible reason (not mentioned in
\cite{Pietr95} and \cite{Rich}) to exclude the lower group from the OSR
systematics. We have in mind spin-flip excitations found in $^{164}$Dy by
D. Frekers at al \cite{Frek}: $B_{\sigma}(M1) = 0.72\ \mu_N^2$ at $E$ = 2.53 
MeV and $B_{\sigma}(M1) = 0.50\ \mu_N^2$ at $E$ = 3.14 MeV. The
spin contribution at $E$ = 3.14 MeV was subtracted in both OSR systematics
\cite{Pietr95,Rich}. In such a way it was found for OSR strength
$B(M1)=3.18\ \mu_N^2$ in \cite{Pietr95} and 
$B(M1)=3.25\ \mu_N^2$ in \cite{Rich}.

\cite{Nest}, page 5, left column: 

"However,
following our results in Fig. 3 the states at 2.4 -- 2.7 MeV give mainly
orbital $M1$ transitions and so should also belong to OSR. They are omitted in
OSR systematics with the lower boundary 2.7 MeV \cite{Pietr95} but taken into account for the lower boundary 2.5 MeV \cite{Rich}".

First of all, the agreement between calculated and experimental data is  
too bad to make so resolute conclusions. Even the general distribution of the $M1$ strength contradicts to the experimental situation:
the strength of the lower group of calculated levels is larger than that of
the higher group! The inference "... mainly orbital ..." seems doubtful
because of the very strong constructive interference, that says about the
huge influence of spin degrees of freedom (independently of the value of 
the contribution of the spin part of the magnetic dipole operator). 
The final statement about the paper \cite{Rich} is simply not true.

\begin{table}[t!] 
\caption{$^{164}$Dy. Contributions of various currents into calculated
excitations. (a), (b),...,(f) -- numbers of pictures in Figs.~\ref{E1}, \ref{E2}, \ref{E3}.}
\begin{ruledtabular}\begin{tabular}{ccc}
 $E$ (MeV)            & (i), (j) & \% \\ 
\hline
\multirow{3}{*}{2.20} & (a), (b) & {~1.75} \\  
                      & (c), (d) & {47.29} \\
                      & (e), (f) & {50.95} \\ 
\hline
\multirow{3}{*}{2.87} & (a), (b) & {31.90} \\
                      & (c), (d) & {53.71} \\
                      & (e), (f) & {14.39} \\
\hline
\multirow{3}{*}{3.59} & (a), (b) & {61.55} \\
                      & (c), (d) & {~7.76} \\
                      & (e), (f) & {30.69} \\
\end{tabular}\end{ruledtabular}\label{tab2}
\end{table}       

The solution of TDHFB equations by WFM method for $^{164}$Dy gives three 
low lying magnetic states with the
following energies and magnetic strengths:
$E_1$ =2.20 MeV, $B_1(M1) =1.76\ \mu_N^2$,
$E_2$ =2.87 MeV, $B_2(M1) =2.24\ \mu_N^2$,
$E_3$ =3.59 MeV, $B_3(M1) =1.56\ \mu_N^2$.
According to WFM results the energy area 2.4 -- 2.7 MeV
is mainly of the spin character. Really, the analysis of Table~\ref{tab2}
allows one to conclude that:
\begin{enumerate}
\item  excitation with $E=2.20$ MeV represents
predominantly ($51\% $) the "complicate" spin scissors (Fig.~\ref{E1}~(e),~(f)) with rather strong 
admixture ($47\% $) of the "simple" spin scissors (Fig.~\ref{E1}~(c),~(d)), 
\item  excitation with $E=2.87$ MeV represents
predominantly ($54\% $) the "simple" spin scissors (Fig.~\ref{E2}~(c),~(d)) with rather big 
admixture ($32\% $) of the conventional scissors (Fig.~\ref{E2}~(a),~(b)),
 \item  excitation with $E=3.59$ MeV represents
predominantly ($62\% $) the conventional scissors (Fig.~\ref{E3}~(a),~(b)) with a rather strong admixture ($31\% $) of the "complicate" spin scissors 
(Fig.~\ref{E3}~(e),~(f)).
\end{enumerate}

As one can see the excitation with E = 2.2 MeV has the pure spin
nature (the mixture of simple and complicate spin scissors). 
On the other hand the excitation
with E = 2.87 MeV has the mixed structure: 68\% of the spin nature (simple
and complicate spin scissors) and 32\% of the orbital nature (conventional
scissors). So, it is natural to expect that in the case of splitting of 
these two excitations the energy interval between them will be filled mainly
by the excitations of the spin nature.

\subsection{Spin scissors or spin-flip?}

\cite{Nest}, page 6, right column:

"so called SSR states are actually
ordinary low-energy non-collective spin-flip excitations which do not need 
for their explanation the scissors-like treatment".

Surprising declaration! Let us remind that different methods of the 
solution of any problem (TDHFB equations in our case) use as a rule 
different "instruments" and different languages. As a matter of fact 
spin scissors and spin-flip
are just different names of the same physical phenomenon.
Really, RPA deals with transitions between various levels (particle-hole
excitations). On the other hand WFM operates with the various phase space
moments of a nucleus (second order moments in our case: quadrupole moments
in coordinate and momentum spaces and an angular moment). Both methods 
produce several 1$^+$ states with quite close values of summed $M1$ 
strengths. Some part of 
these strengths appears due to the spin-orbital potential. Obviously,
from the microscopic (RPA) point of view it is the transition between 
spin-orbital partners (i.e. spin-flip). In the macroscopic approach (WFM) it 
is interpreted as the counter-rotation of spin up nucleons with respect of 
spin down nucleons (i.e. spin scissors) because here it is generated by 
counter-oscillations of their orbital angular momenta.

By the way, transitions between spin-orbital 
partners represent only the particular case among all possible  transitions. 
Let us remember for example, that from the 
microscopic point of view the conventional (orbital) scissors also are 
produced by transitions between some levels inside of one major shell
(i.e. $\Delta$N=0). 
The scissors-like nature of the considered excitation
can be revealed by constructing the picture of currents or calculating 
the angular momenta of all four constituents (spin up and spin 
down protons and neutrons) of the excited nucleus.
The lines of currents are already produced in the WFM approach \cite{BM3S}
-- they are reproduced here in Figs.~\ref{E1}, \ref{E2}, \ref{E3}. The angular momenta  belong to the set of variables of the method. 

A special comment is required for the
situation in Fig.~\ref{E1}~(a),~(b) where both currents turn in the same 
direction leading to the impression that the total angular momentum is not 
zero as it should be.  We remark, however, that the total angular momentum 
zero is well conserved, the counter-rotation 
being performed by the motion of the spins.

Having RPA wave function one
must not meet any difficulties to do the same: to draw the lines of 
currents and to calculate the respective angular momenta.

\subsection{M1 strength in $^{232}$Th}

\cite{Nest}, page 8, left column:

"there is a remarkable agreement between the distribution of the total $M1$ strength and the experimental data" 

This statement looks non 
convincing, because the summed $M1$ strength of the higher group of levels is 
larger than that of the lower group in contradiction with the experimental
distribution. After it their final inference that 

"two level groups are
explained not by separation of SSR and  OSR modes (as was suggested by WFM)
but rather by a fine structure of the OSR alone" 

looks rather strange.

\begin{figure}[h!]
\centering{\includegraphics[width=\columnwidth]{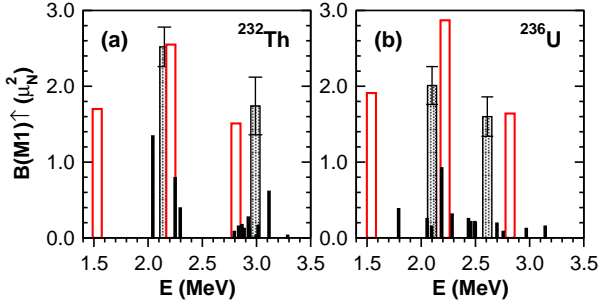}}
\caption{The centroids of experimentally observed spectra of $1^+$ 
excitations in $^{232}$Th (a) and $^{236}$U (b) (black rectangles with error 
bars) are compared with the results of WFM calculations (red 
rectangles).}\label{f5}
\end{figure}


It will be useful to remind here WFM results. The solution of TDHFB 
equations for $^{232}$Th gives three low lying magnetic states with the
following energies and magnetic strengths:
$E_1$ =1.53 MeV, $B_1(M1) =1.7\ \mu_N^2$,
$E_2$ =2.21 MeV, $B_2(M1) =2.55\ \mu_N^2$,
$E_3$ =2.81 MeV, $B_3(M1) =1.51\ \mu_N^2$.
As it is seen from Fig. \ref{f5} the second and third states reproduce very well
the values of energy centroids and summed $B(M1)$ of the lower and higher groups of 1$^+$ excitations observed in $^{232}$Th. 
The summed $M1$ strength of these two levels $B(M1) = 4.07\ \mu_N^2$ and 
their energy centroid $E$ = 2.43 MeV practically coincide with the 
respective
experimental data $B_{exp}(M1) = 4.26\ \mu_N^2$ and $E_{exp}$ = 2.49 MeV.
The lowest calculated level $E_1$ is the prediction. The picture of currents 
in $^{232}$Th is indistinguishable from that of $^{164}$Dy. The internal structure of three calculated excitations also is quite close to that of 
$^{164}$Dy (see Table~\ref{tab3}). The analysis of this table allows one to conclude that:
\begin{enumerate}
\item  excitation with $E=1.53$ MeV is the admixture
of "simple" (Fig.~\ref{E1}~(c),~(d)) and "complicate" 
(Fig.~\ref{E1}~(e),~(f)) spin scissors
with their practically equal contributions,
\item  excitation with $E=2.21$ MeV represents
predominantly ($48\% $) the "simple" spin scissors (Fig.~\ref{E2}~(c),~(d)) with rather big 
admixture ($36\% $) of the conventional scissors (Fig.~\ref{E2}~(a),~(b)),
 \item  excitation with $E=2.81$ MeV represents
predominantly ($60\% $) the conventional scissors (Fig.~\ref{E3}~(a),~(b)) with a rather strong admixture ($31\% $) of the "complicate" spin scissors 
(Fig.~\ref{E3}~(e),~(f)).
\end{enumerate}

\begin{table}[h!] 
\caption{$^{232}$Th. Contributions of various currents into calculated
excitations. (a), (b),...,(f) -- numbers of pictures in Figs.~\ref{E1}, \ref{E2}, \ref{E3}.
}
\begin{ruledtabular}\begin{tabular}{ccc}
 $E$ (MeV)            & (i), (j) & \% \\ 
\hline
\multirow{3}{*}{1.53} & (a), (b) & {~2.73} \\  
                      & (c), (d) & {48.95} \\
                      & (e), (f) & {48.32} \\ 
\hline
\multirow{3}{*}{2.21} & (a), (b) & {36.41} \\
                      & (c), (d) & {48.41} \\
                      & (e), (f) & {15.18} \\
\hline
\multirow{3}{*}{2.81} & (a), (b) & {60.29} \\
                      & (c), (d) & {~9.20} \\
                      & (e), (f) & {30.51} \\
\end{tabular}\end{ruledtabular}\label{tab3}
\end{table}    

\section{Conclusion}

We have explained that there is no contradiction between
different names of low 
lying 1$^+$ states: spin scissors in WFM and spin-flip in QRPA. Spin-flip 
appears in the microscopic approach which deals with transitions between various quantum-mechanical states (levels). To see what kind of macroscopic
motion (rotation or vibration) is hidden behind of this
quantum picture, it is necessary to calculate the 
currents or (and) angular moments of spin up and down protons and neutrons.

Non collective  character is the old puzzle of the scissors mode.
Compromise 
solution of this "problem" was suggested in \cite{Heyd}: scissors mode is
"weakly collective, but strong on the single-particle scale". It is the very 
strong and attractive feature (property) of the WFM method, that operating
by collective variables it is able to describe the "weakly collective" 
phenomena.

\cite{Nest}, page 9, left column: 

"So heavy nuclei are generally not suitable
to exhibit LMSF (Low-Moment Spin-Flip)". 

This inference contradicts to our analysis of $^{232}$Th, especially to the prediction of the pure spin scissors (see above).

\cite{Nest}, page 9, left column: 

"Our calculations show that the lowest $1^+$
states in $^{160,162,164}$Dy are indeed of the spin-flip character. However
they are located at $E \le 2.4$ MeV, i.e. below the observed states."

Probably there are some problems with the used method or with the chosen 
forces. Contrary to the situation in \cite{Nest}, the results of WFM
calculations for $^{164}$Dy and $^{232}$Th are in the excellent agreement
with experimental data.

\cite{Nest}, page 9, right column: 

"So, by our opinion, the available 
experimental data do not confirm the existence of SSR".

Rather strange declaration... To be exact, it is the calculations of \cite{Nest} do not confirm the SSR existence. And not surprisingly, because
they don't agree with experimental data!

\cite{Nest}, page 9, right column: 

"The WFM {\it scissors-like} treatment of
SSR requires the nuclear deformation."

It is not true. We have shown analytically in \cite{BaSc} that a deformation 
is necessary for an existence of a conventional (orbital) scissors. But we
never said it about the spin scissors!

\end{document}